\documentclass[11pt,a4paper,twoside]{article}

\usepackage{epsfig}
\usepackage[latin1]{inputenc}
\def\be{\begin{equation}}
\def\ee{\end{equation}}
\def\ba{\begin{array}}
\def\bacc{\begin{array} {cc}}
\def\ea{\end{array}}
\def\bea{\begin{eqnarray}}
\def\eea{\end{eqnarray}}
\def\bd{\begin{displaymath}}
\def\ed{\end{displaymath}}

\def\a{\alpha}
\def\b{\beta}

\def\ob{\overline{\omega}}

\def\Box{ {\,\lower 0.9pt\vbox{\hrule\hbox{\vrule height0.2cm \hskip 0.2cm
\vrule height 0.2cm }\hrule}\,}}

\begin{document}

\begin{center}

{\Large\bf Gauge Fields, Fermions and \\ Mass Gaps in 6D Brane
Worlds }

\vspace{1cm}

{\large S. L. Parameswaran$^a$\footnote{Email: param@sissa.it}, S.
Randjbar-Daemi$^b$\footnote{Email: seif@ictp.trieste.it} and
A. Salvio$^a$\footnote{Email: salvio@sissa.it}}\\

\vspace{.6cm}

{\it {$^a$ International School for Advanced Studies,\\
Via Beirut 2-4, 34014 Trieste, Italy}}

\vspace{.4cm}

{\it {$^b$ International Center for Theoretical Physics, \\Strada
Costiera 11, 34014 Trieste, Italy}}

\end{center}

\vspace{1cm}

\begin{abstract}

We study fluctuations about axisymmetric warped brane solutions in
6D minimal gauged supergravity. Much of our analysis is general and
could be applied to other scenarios. We focus on bulk sectors that
could give rise to Standard Model gauge fields and charged matter.
We reduce the dynamics to Schroedinger type equations plus physical
boundary conditions, and obtain exact solutions for the Kaluza-Klein
wave functions and discrete mass spectra.  The power-law warping, as
opposed to exponential in 5D, means that zero mode wave functions
can be peaked on negative tension branes, but only at the price of
localizing the whole Kaluza-Klein tower there. However, remarkably,
the codimension two defects allow the Kaluza-Klein mass gap to
remain finite even in the infinite volume limit.  In principle, not
only gravity, but Standard Model fields could `feel' the extent of
large extra dimensions, and still be described by an effective 4D
theory.

\end{abstract}

\newpage

\tableofcontents

\newpage

\section{Introduction} \label{intro}

Minimal 6D gauged supergravity and its solutions have received much interest
recently for several reasons.  From the top down, the theory shares many
features in common with 10D supergravity, whilst remaining relatively simple,
and so it can be used as a toy model for 10D string theory compactifications
\cite{toykklt}.  From the bottom up, it provides a context in which to extend
the well-trodden path of 5D brane world models to codimension two.  Moreover, in
6D models, Supersymmetric Large Extra Dimensions (SLED) have shown some promise
in addressing the two overshadowing fine-tuning problems of fundamental physics:
 the Gauge Hierarchy and the Cosmological Constant Problems \cite{sled}.

In terms of the phenomenological study of brane worlds, one should ask what are
the qualitative differences between 5D and 6D models.  For example, in five
dimensional Randall-Sundrum (RS) models \cite{RS} the warping of 4D spacetime
slices is exponentially dependent on the proper radius of the extra dimension,
whereas in the six dimensional models developed in Refs.~\cite{GGP,
warpedbranes, Burgess:2004dh} it is only power law dependent.  Moreover, the
singularities sourced by the branes are distinct, the codimension one case being
a jump and the codimension two case being conical.

5D RS models with large (or infinite) extra dimension and the Standard Model
(SM) confined to the brane were developed to explain the hierarchy in the Planck
and Electroweak scales.  Although the mass gap in the Kaluza-Klein spectrum goes
to zero as usual in the infinite volume limit, 4D physics is retrieved thanks to
the warp factor's localization of the zero mode graviton - and exponential
suppression of higher modes - close to the brane with positive tension.
Subsequently it was found that the step singularities in the geometry could also
localize bulk fermions \cite{5Dfermions}, in much the same way as previously
achieved with scalar fields and kink topological defects \cite{kink}.  Models
were then developed in which SM fields all\footnote{Although the Higgs field
should be confined to the brane in order not to lose the gauge hierarchy.}
originate from the bulk as localized degrees of freedom \cite{5DSM}.

A study of warped brane worlds in 6D supergravity was intitiated in
\cite{GGP, warpedbranes, Burgess:2004dh}, where the focus was on the
background solutions. As a result of supersymmetry
\cite{Nishino:1984gk}, and anomaly cancellation \cite{seifanomaly},
the 6D theory is highly constrained, and moreover fairly general
classes of solutions have been obtained.  In \cite{GGP}, a general
solution with 4D Poincar\'e and 2D axial symmetry was derived, and
it was shown that warping always leads to conical (or worse)
singularities in the internal manifold, which can be naturally
interpreted as 3-branes sources. It is certainly interesting to go
beyond these background solutions, and study the dynamics of their
fluctuations.  The final objective would be to obtain the effective
theory describing 4D physics, and an understanding of when this
effective theory is valid.

Although in these constructions the SM is usually put in by hand, envisioned on
a 3-brane source, the bulk theory is potentially rich enough to contain the SM
gauge and matter fields.  As hinted above, should the SM arise as Kaluza-Klein
zero modes of bulk fields, there are two ways to hide the heavy modes and
recover 4D physics.  They may have a large mass gap, and thus be unattainable at
the energy scales thus-far encountered in our observed universe.  Or they may be
light but very weakly coupled to the massless modes, for example if the massless
modes are peaked near to the brane, and the massive modes are not.  In any case,
whether or not one expects the bulk to give rise to the SM, one should study its
degrees of freedom and determine under which conditions they are observable or
out of sight.  This could also prove useful for a deeper understanding of the
self-tuning mechanism of SLED, and its quantum corrections.

A complete study of the linear perturbations is a very complicated problem,
involving questions of gauge-fixings and a highly coupled system of dynamical
equations.  Some partial results have been obtained for the scalar perturbations
in \cite{leepapa}.  In this paper we consider sectors within which SM gauge and
charged matter fields might be found.  By some fortune, these also happen to be
two of the least complicated ones.  Much of our discussion is general, and could
easily be applied or extended to other 6D models with axial symmetry.  We follow
the usual Kaluza-Klein procedure, and reduce the equations of motion to an
equivalent non-relativistic quantum mechanics problem, which we are able to
solve exactly.  We consider carefully the boundary conditions that the physical
modes must satisfy, and from these derive the wave function profiles and
complete discrete mass spectra.

Our exact solutions enable us to analyze in detail the effects of the power-law
warping and conical defects that arise in 6D brane worlds.
We find that the warping cannot give rise to zero modes peaked at the brane,
without also leading to peaked profiles for the entire Kaluza-Klein tower.  On
the other hand, the conical defects do break another standard lore of the
classical Kaluza-Klein theory.  Remarkably, even if the volume of the internal
manifold goes to infinity, the mass gap does not necessarily go to zero.  This
decoupling between the mass gap and volume means that in principle SM fields, in
addition to gravity, could `feel' the extent of large extra dimensions, whilst
still being accurately described by a 4D effective field theory.

The paper is organised as follows.  We begin in Section \ref{6dtheory} by
reviewing the 6D supergravity theory, and its axially symmetric warped brane
world solutions.  Then, in Section \ref{gaugefields} we analyze the gauge field
fluctuations, deriving the wave functions and masses of the Kaluza-Klein
spectrum.  A similar analysis is presented in Section \ref{fermion} for the
fermions.  Section \ref{LargeMV} discusses the physical implications of the
results found, and in particular whether they can be naturally applied to the
Large Extra Dimension scenario.  Finally we end in Section \ref{conclusions}
with some conclusions and future directions.

In the appendices we give some results that are useful for the detailed
calculations.  Appendix \ref{conventions} sets our conventions, and Appendix
\ref{deltafn} explains how the conical defects manifest themselves in the metric
ansatz.  In Appendix \ref{boundcond} we show in detail how the boundary
conditions are applied to obtain a discrete mass spectrum, and we give the
complete fermionic mass spectrum thus derived in Appendix \ref{fermspectrum}.

\section{6D Supergravity and Axisymmetric Solutions} \label{6dtheory}
\setcounter{equation}{0}

Let us begin by presenting 6D ${\cal N}=1$ gauged
supergravity, and its warped brane world solutions, whose
fluctuations we will then study.  The field content consists of a
supergravity-tensor multiplet ($G_{MN}$, $B_{MN}$, $\sigma$,
$\psi_M$, $\chi$)  - with metric, anti-symmetric Kalb-Ramond field
(with field strength $H_{MNP}$), dilaton, gravitino and dilatino -
coupled to a number of gauge multiplets (${\cal A}_M$, $\lambda$) - with
gauge potentials and gauginos - and a number of hypermultiplets
($\Phi$, $\Psi$) - with hyperscalars and hyperinos.  The fermions
are all Weyl spinors, satisfying $\Gamma_7 \psi_M = \psi_M$,
$\Gamma_7 \chi = - \chi$, $\Gamma_7 \lambda = \lambda$ and
$\Gamma_7 \Psi = - \Psi$, and in general the theory has anomalies.
However, for certain gauge groups and hypermultiplet
representations these anomalies can be cancelled via a
Green-Schwarz mechanism \cite{seifanomaly, anomaly}.  We will consider a
general matter content, with gauge group of the form
$\mathcal{G}=\tilde{\mathcal{G}}\times U(1)_R$, where $U(1)_R$ is an R-symmetry,
and $\tilde{\mathcal{G}}$ is in general a product of simple groups.  For example
we could take the anomaly free group
$\mathcal{G}=E_6 \times E_7 \times U(1)_R$, under which the
fermions are charged as follows: $\psi_M \sim (1,1)_1$, $\chi \sim
(1,1)_1$, $\lambda \sim (78,1)_1 + (1,133)_1 + (1,1)_1$, $\Psi \sim (1,
912)_0$ \cite{seifanomaly}.

The bosonic action takes the form\footnote{We give conventions and
notation in Appendix \ref{conventions}.  For fermionic terms see
\cite{Nishino:1984gk}.} \cite{Nishino:1984gk}
\bea S_B &=&\int d^6 X
\sqrt{-G}\left[\frac{1}{\kappa^2}R-\frac{1}{4}\partial_M\sigma\partial^M\sigma
-\frac{1}{4}e^{\kappa\sigma/2}Tr\left( \frac{1}{{\tilde g}^2}{\tilde F}^2 +
\frac{1}{g_1^2}F_1^2 \right)
\right.\nonumber \\
&&\qquad \qquad \qquad
-\frac{1}{12}e^{\kappa\sigma}H_3^2 \left.-G_{\a
\b}(\Phi)D_M\Phi^{\a}D^M\Phi^{\beta} -\frac{8}{\kappa^4}e^{-\kappa
\sigma/2}v(\Phi)\right],
\label{SB} \eea
where $\kappa$ represents the 6D Planck scale, $\tilde{g}$ and $g_1$
are the $\tilde{\mathcal G}$ and $U(1)_R$ gauge coupling constants
respectively\footnote{In general, since $\tilde{\mathcal G}$ consists of several
simple factors, $\tilde{g}$ represents a collection of independent gauge
couplings.} and $\tilde{F}_{MN}$ and $F_{1\,MN}$  the corresponding
field strengths. $G_{\a\b}(\Phi)$ is the metric on the target
manifold of the hyperscalars, and here the index $\a$ runs over all
the hyperscalars.  The dependence of the scalar potential on
$\Phi^{\a}$ is such that its minimum is at $\Phi^{\a} = 0$, where it takes a
positive-definite value, $v(0)
= g_1^2$, due to the R-symmetry gauging \cite{NS2, dyonicstring}.  We therefore
fix $\Phi^{\a} =0$.  The remaining equations of motion (EOM) are
\bea &&\frac{1}{\kappa^2}R_{MN}= \frac{1}{4}\partial_{M}\sigma\partial_N\sigma
+
\frac{1}{2}e^{\kappa\sigma/2}\left[\frac{1}{\tilde{g}^2}Tr\left(\tilde{F}_{MP}
\tilde{F}^{\,\,\,\,\,P}_{N}\right)
+\frac{1}{g_1^2}Tr\left(F_{1\,MP}F^{\,\,\,\,\,\,\,P}_{1\,N}\right)\right]
\nonumber \\
&& \phantom{00000000000} + \frac{1}{4} e^{\kappa\sigma} H_{MPQ} H_N^{\,\,\, PQ}
-\frac{1}{4\kappa}G_{MN}\Box\sigma, \nonumber \\
 &&\frac{1}{\kappa}\Box\sigma=\frac{1}{4}e^{\kappa\sigma/2}\left[\frac{1}{\tilde
{g}^2}Tr\tilde{F}^2
+\frac{1}{g_1^2}TrF_1^2\right]+ \frac{1}{6} e^{\kappa\sigma} H_{MNP}H^{MNP}
-\frac{8g_1^2}{\kappa^4}e^{-\kappa\sigma/2},\nonumber \\
&& D_M\left(e^{\kappa\sigma/2}\tilde{F}^{MN}\right)=\frac{\kappa}{2}
e^{\kappa\sigma} H^{NPQ} {\tilde F}_{PQ},\quad
D_M\left(e^{\kappa\sigma/2}F_1^{MN}\right)=\frac{\kappa}{2} e^{\kappa\sigma}
H^{NPQ} F_{1\, PQ}, \nonumber \\
&& D_M\left(e^{\kappa\sigma}H^{MNP}\right)=0,
\label{EOM}\eea
where $D_M$ is the gauge and Lorentz covariant derivative.

We will consider a general class of warped solutions with 4D
Poincar\'e symmetry, and axial symmetry in the transverse
dimensions:
\bea ds^2=G_{MN}dX^M dX^N&=&e^{A(\rho)}\eta_{\mu
\nu}dx^{\mu}dx^{\nu}+d\rho^2+e^{B(\rho)}d\varphi^2,\nonumber\\
\mathcal{A}&=&\mathcal{A}_{\varphi}(\rho) Q d\varphi,\nonumber \\
\sigma &=&\sigma(\rho),\nonumber \\
H_{MNP} &=& 0,
\label{axisymmetric} \eea
with $0\leq \rho \leq \overline{\rho}$ and $0\leq \varphi<2\pi.$ Here
$\mu,\nu=0,1,2,3$, $\mathcal{A}$ is a gauge field, and $Q$ is a generator of a
$U(1)$ subgroup of a simple factor of $\mathcal{G}$, satisfying
$Tr\left(Q^2\right)=1$.

In the following we shall also use the radial coordinate defined by
\be u(\rho)\equiv \int_0^{\rho}d\rho' e^{-A(\rho')/2},
\label{u}\ee
whose range is $0\leq u \leq \overline{u}$. In this frame the metric reads
\be ds^2=e^{A(u)}\left(\eta_{\mu \nu}dx^{\mu}dx^{\nu}+du^2\right)
+e^{B(u)}d\varphi^2 \, .\ee

Given the above ansatz, the general solution to the equations of
motion (\ref{EOM}) has been found by Gibbons, G\"uven and Pope (GGP) in
\cite{GGP}\footnote{Relaxing
the condition of 4D Poincar\'e symmetry to that of only 4D maximal
symmetry should allow more general solutions to be found
\cite{andrew}.  More general solutions also exist breaking axial
symmetry \cite{nonaxial}, or having nontrivial VEVs for the
hyperscalars \cite{superswirl}.}.  Although much of our formalism for the
perturbation analysis can be applied to the general ansatz
(\ref{axisymmetric}), we will focus on a subset of this general solution, namely
that
which contains singularities no worse than conical. Thus, in
addition to the ansatz (\ref{axisymmetric}), we impose the
following asymptotic behaviour for the metric:
$$e^A\stackrel{\rho\rightarrow0}{\rightarrow}constant \neq 0,\quad
e^A\stackrel{\rho\rightarrow \overline{\rho}}{\rightarrow}
constant \neq 0,$$ and
\be \quad
e^B\stackrel{\rho\rightarrow0}{\rightarrow}\left(1-\delta/2\pi\right)^2\rho^2,
\quad e^B\stackrel{\rho\rightarrow \overline{\rho}}{\rightarrow}
\left(1-\overline{\delta}/2\pi\right)^2\left(\overline{\rho}-\rho\right)^2,
\label{assumeeB}\ee
that is we assume conical defects with deficit angle
$\delta$ at $\rho=0$ and $\overline{\delta}$ at
$\rho=\overline{\rho}$, at which points the Ricci scalar contains
delta-functions (see Appendix \ref{deltafn}).  These can be
interpreted as 3-brane sources with tensions $T=2\delta/\kappa^2$
and
$\overline{T}=2\overline{\delta}/\kappa^2$ \cite{chenlutyponton}.

The explicit conical-GGP solution is then\footnote{The coordinate
$u$ is related to the coordinate $r$ in \cite{GGP} by
$r=r_0\cot(u/r_0)$.} \cite{GGP}:
\bea e^A&=&e^{\kappa \sigma/2}=\sqrt{\frac{f_1}{f_0}}, \quad e^B=\alpha^2
e^A\frac{r_0^2\cot^2(u/r_0)}{f_1^2},\nonumber\\
\mathcal{A}&=&-\frac{4\alpha g}{q\kappa f_1}\, Q\,
d\varphi,\label{GGPsolution}\eea
where $\alpha$ is a real number, which we can take to be positive without loss
of generality, $q$ is a real number and
$g$ is the gauge coupling constant corresponding to the
background gauge field.  For example, if ${\mathcal A}$ lies in $U(1)_R$, then
$g=g_1$.  Also,
\be f_0\equiv 1+\cot^2\left(\frac{u}{r_0}\right), \quad f_1 \equiv
1+\frac{r_0^2}{r_1^2}\cot^2\left(\frac{u}{r_0}\right), \ee
with $r_0^2\equiv \kappa ^2/(2g_1^2)$, $r_1^2\equiv 8/q^2$.  Moreover,
$\overline{u}\equiv \pi r_0/2$ and the deficit
angles are
\bea \delta&=&2\pi \left(1-\alpha \frac{r_1^2}{r_0^2}\right)\, , \\
\overline{\delta}&=&2\pi\left(1-\alpha\right) \, . \eea
Notice then that a non-trivial warping enforces the presence of a 3-brane
source.  On the other hand, the parameter $\alpha$ is not fixed by the EOM and
it represents a
modulus.

The expression for the gauge field background in eq. (\ref{GGPsolution}) is
well-defined in the limit $u\rightarrow 0$, but not as $u \rightarrow
\overline{u}$.  We should therefore use a different patch to describe the
$u={\overline{u}}$ brane, and this must be related to the patch including the
$u=0$ brane by a single-valued gauge transformation.  This leads to a Dirac
quantization condition, which for a field interacting with
$\mathcal{A}$ through a charge $e$ gives
\be e \, \frac{4\alpha g}{\kappa q}= e \,\alpha \frac{r_1}{r_0} \frac{g}{g_1} =
N \, , \label{DiracQ}\ee
where $N$ is an integer.  The charge $e$ can be computed once we have selected
the background gauge group, since it is an eigenvalue of the generator $Q$.
Finally, note that the internal manifold corresponding to solution
(\ref{GGPsolution}) has an $S^2$ topology (its Euler number
equals 2).

\lineskip
\lineskip
\lineskip
\lineskip

We end this section by considering the various parameters in the
model, and the phenomenological constraints which can arise when
we give it a brane world interpretation.  There are three free parameters in the
6D theory, which can be
taken to be the gauge coupling ${\tilde g}$, and two out of the following three
parameters: the 6D
Planck scale, $\kappa$, the gauge coupling $g_1$ and the
length-scale $r_0 = \kappa/\sqrt{2} g_1$. In the solution there are
two free parameters, $r_1$ (or $q$) and $\alpha$. However, one
combination of all these parameters is constrained by the
quantization condition (\ref{DiracQ}).

The relation between the 6D Planck scale $\kappa$ and our observed
4D Planck scale $\kappa_4$ is
\be \frac{1}{\kappa^2}V_2=\frac{1}{\kappa_4^2}, \label{k4}\ee
where the volume $V_2$ is given by
\be V_2=\int d^2y \sqrt{-G}e^{-A}=2\pi \int du e^{(3A+B)/2}. \ee
For solution (\ref{GGPsolution}) we have
\be V_2=4\pi \alpha \left(\frac{r_0}{2}\right)^2. \label{V2}\ee
Notice that this volume does not depend on $r_1$, and so we can keep it fixed
whilst varying the warp factor, namely $e^A$ in (\ref{GGPsolution}).  Moreover
from (\ref{k4}) a
phenomenological constraint follows between the bulk couplings and
the brane tensions, which can be written:
\be \frac{g_1}{\sqrt{\alpha}}=\sqrt{\frac{\pi}{2}}\kappa_4. \ee
This implies that $g_1/\sqrt{\a}$ is very small, of the order of the
Planck length.

Now let us embed the Arkani Hamed-Dimopoulos-Dvali (ADD) scenario
\cite{ADD} into the present model, in order to try to explain the large
hierarchy between the Electroweak scale and the Planck scale via
the size of the extra dimensions.  Thus identifying the 6D
fundamental scale with the Electroweak scale $\kappa\sim TeV^{-2}$
and constraining the observed 4D Planck scale $\kappa_4^2\sim
10^{-30}TeV^{-2}$; the above relation translates to\footnote{We use the
following conversion relation: $(TeV)^{-1} \sim 10^{-16}mm$.}:
\be \sqrt{\alpha}r_0\sim 0.1 mm.\label{submillim}\ee
Here, the large extra dimensions corresponds to tuning the bulk
gauge coupling and brane tensions.  However, we can also observe
that (\ref{submillim}) fixes just one parameter among $\alpha$,
$r_0$ and $r_1$ and we still have two independent parameters even
if we require large extra dimensions. Later we will see that this
novel feature proves to have interesting consequences for the mass
spectrum of fluctuations.

\section{Gauge Fields}\label{gaugefields}
\setcounter{equation}{0}

Having established the brane world solution and its properties, we
are now ready to examine the fluctuations about this background,
which will represent the physical fields in our model.  In this
section our focus will be on the gauge field fluctuations.

Normalizable gauge field zero modes in axially symmetric codimension
two branes are known to exist \cite{Giovannini:2001vt,
Giovannini:2002mk, Randjbar-Daemi:2002pq}. However, in these known
examples there is no mass gap between the zero and non-zero modes
 which renders an effective
4-dimensional description somewhat problematic, especially in non-Abelian case
\cite{Randjbar-Daemi:2003qd}. In contrast to this
 for the axisymmetric solutions studied in this paper the presence of a mass gap
will be automatic
due to the compactness of the transverse space. In this section we
shall give the full spectrum of zero and non-zero modes.

Given the symmetries of the problem, we can expect that the gauge
fields in the low energy effective theory belong to ${\cal H}
\times U(1)_{KK}$, where ${\cal H}$ is the unbroken subgroup of ${\cal G}$
that commutes with the $U(1) \subset {\cal G}$ in which the
monopole lies\footnote{Due to the Chern-Simons coupling in supergravity, the
$U(1)$ gauge field in the direction of the monopole eats the axion arising from
the Kalb-Ramond field and acquires a mass \cite{seifanomaly, toykklt}.}.  The
$U(1)_{KK}$ arises from the vector
fluctuations of the metric, due to the axial symmetry of the
internal manifold, and is promoted to $SU(2)_{KK}$ in the sphere
limit of the background.

The non-Abelian sector of ${\cal H}$ may be rich enough to contain
the SM gauge group.  For example, consider the
anomaly free model of \cite{seifanomaly}, with gauge group ${\cal G} =
E_6 \times E_7 \times U(1)_R$, and the monopole background in
$E_6$.  The surviving gauge group, $SO(10) \times E_7 \times
U(1)_R \times U(1)_{KK}$, then contains the Grand Unified Group
$SO(10)$, and the model also includes charged matter in the fundamental of
$SO(10)$.  Therefore our present interest will be in the fields belonging to
various representations of ${\cal H}$.  Specifically, we will consider gauge
field fluctuations orthogonal to the monopole background.  For the case ${\cal
G}=E_6 \times E_7 \times U(1)_R$, the gauge field sectors that are covered by
our analysis are given in Table \ref{T:summary}.

\subsection{Kaluza-Klein Modes}

Using the
background solution in the 6D action (\ref{SB}), we can identify the
bilinear action for the fluctuations.  This step requires some
care, because to study the physical spectrum we must first
remove the gauge freedoms in the action due to 6D diffeomorphisms
and gauge transformations. The problem has been studied in a general
context in \cite{Randjbar-Daemi:2002pq}, where the authors choose
a light-cone gauge fixing.

In the light-cone gauge, the action for the gauge field
fluctuations, orthogonal to the monopole background, at the
bilinear level reads \cite{Randjbar-Daemi:2002pq}
\be S_G(V,V)\equiv -\int d^6X
\sqrt{-G}\,\frac{1}{2}e^{\phi}\left(\partial_{\mu}V_j\partial^{\mu}V^j
+e^{-A}\partial_{\rho}V_j\partial_{\rho}V_j
+D_{\varphi}V_jD^{\varphi}V^j\right), \label{SVV}\ee
where $V_j$ is the gauge field fluctuation in the light cone gauge
(j=1,2) and all the indices in (\ref{SVV}) are raised and lowered
with the $\rho$-dependent metric $G_{MN}$ given in
(\ref{axisymmetric}).  Indeed, here and below $G_{MN}$ represents the background
metric.  We have multiplied the formula of
\cite{Randjbar-Daemi:2002pq} by an overall $e^\phi$, with $ \phi \equiv
\kappa\sigma/2$ and $\sigma$ in the background, due to the presence of the
dilaton in our theory\footnote{That the dilaton invokes only this simple change
with respect to Ref.~\cite{Randjbar-Daemi:2002pq} can be seen as follows.
First, notice that since we are considering fluctuations orthogonal to the
$U(1)$ background, there are no mixings with other sectors, and the bilinear
action is simply $S_G = -1/4 \int d^6X \sqrt{- G }
e^{\phi}G^{MN}G^{PQ}\left(F_{MP}F_{NQ}\right)^{(2)}$.  We emphasise that
$G_{MN}$ and $\phi$ now signify the background fields.  Also, $()^{(2)}$
indicates the bilinear part in the fluctuations.  Next, make the change of
coordinates, $d\rho = e^{-\phi/2} d{\tilde\rho}$, and rewrite the background
metric in (\ref{axisymmetric}) as $ds^2 = e^{-\phi} \left( e^{{\tilde A}}
\eta_{\mu\nu} dx^{\mu} dx^{\nu} + d{\tilde\rho}^2 + e^{{\tilde B}} d\varphi^2
\right)$, with ${\tilde A} \equiv A+\phi $ and ${\tilde B} \equiv B+\phi$.  In
this way, the bilinear action, $S_G$, reduces to exactly the same form as that
of Ref.~\cite{Randjbar-Daemi:2002pq}, and we can proceed as they do to transform
into light-cone coordinates, fix the light-cone gauge, and eliminate redundant
degrees of freedom using their equations of motion.}.
Notice that since we are looking at the sector
orthogonal to the monopole background, the Chern-Simons term does
not contribute, and the action takes a simple form.

In general, the covariant derivative $D_{\varphi}V_j$ includes the
gauge field background
\be
D_{\varphi}V_j=\partial_{\varphi}V_j+ie_V\mathcal{A}_{\varphi}V_j,\ee
where again the charge $e_V$ can be computed using group theory once the gauge
group $\tilde{\cal G}$ is chosen. The value $e_V=0$ corresponds to the
gauge fields in the 4D low energy effective theory.  However, since we
can do so without much expense, we keep a generic value of $e_V$.  Those
fluctuations with $e_V\neq 0$ corresponds to vector fields in a
non-trivial representation of the 4D effective theory gauge group.
The Dirac quantization condition (\ref{DiracQ}) then gives $
e_V \, 4\alpha g/(\kappa q)=N_V$, where $N_V$ is an integer.

Next we perform a Kaluza-Klein expansion of the 6D fields.  Since
our internal space is topologically $S^2$, we require gauge fields
to be periodic functions of $\varphi$:
\be V_j(X)=\sum_m V_{jm}(x)f_m(\rho)e^{im\varphi},\label{FourierV}
\ee
where $m$ is an integer.

If we put (\ref{FourierV}) in (\ref{SVV}) we obtain kinetic terms
 for the 4D effective fields proportional to
\be  \int d^4 x \sum_m \eta^{\mu
\nu}\partial_{\mu}V_{jm}^{\dagger}\partial_{\nu}V_{jm}\int d\rho
e^{\phi+B/2}|f_m|^2. \ee
Therefore physical fluctuations, having a finite kinetic energy,
must satisfy the following normalizability condition (NC):
\be \int du |\psi|^2 <\infty, \label{NCV}\ee
where
\be \psi = e^{(2\phi+A+B)/4}f_m. \ee
The quantity $\left|\psi\right|^2$ represents the probability
density of finding a gauge field in $[u,u+du]$.

In fact, this is not the only condition that the physical fields
must satisfy. If we want to derive the EOM from (\ref{SVV})
through an action principle we have to impose\footnote{Actually we
impose that for every pair of fields $V_j$ and $V'_j$ the
condition $\int
d^6X\partial_{M}\left(\sqrt{-G}e^{\phi-A}V_j{\cal D}^{M}V'_j\right)=0$
is satisfied but in (\ref{HCV}) the prime is understood.} the
following boundary condition
\cite{Nicolai:1984jg,Gibbons:1986wg,Kehagias:1999aa}
\be \int
d^6X\partial_{M}\left(\sqrt{-G}e^{\phi-A}V_j{\cal D}^{M}V_j\right)=0 \,
,\label{HCV}\ee
where ${\cal D}_M$ is the gauge covariant derivative.  Equation (\ref{HCV})
represents conservation of current
$J_M=e^{\phi-A}V_j{\cal D}_M V_j$.  Moreover, since the fields are
periodic functions of $\varphi$, (\ref{HCV}) becomes
\be \left[\sqrt{-G}\,e^{\phi-A}\,V_j \partial_{\rho} V_j
\right]_0^{\overline{\rho}}=0.\label{HC2}\ee

The EOM can then be derived as:
\be \sqrt{-G}e^{\phi-2A}\eta^{\mu \nu}\partial_{\mu}\partial_{\nu}V_j=
-\partial_{\rho}\left(\sqrt{-G}
e^{\phi-A}\partial_{\rho}V_j\right)-\sqrt{-G}e^{\phi-A-B}D_{\varphi}^2V_j.\label
{EOMV}\ee
By inserting (\ref{FourierV}) in (\ref{EOMV}) we obtain
\be
-\frac{e^{-\phi+2A}}{\sqrt{-G}}\partial_{\rho}\left(\sqrt{-G}e^{\phi-A}\partial_
{\rho}f_m\right)
+e^{A-B}\left(m+e_V\mathcal{A}_{\varphi}\right)^2f_m=M_{V,m}^2f_m,\label{EOMm}
\ee
where $M_{V,m}^2$ are the eigenvalues of $\eta^{\mu
\nu}\partial_{\mu}\partial_{\nu}$.

At this stage, we can already identify the massless fluctuation
that is expected from symmetry arguments.  For $e_V=0$, when $m=0$, a constant
$f_0$ is a solution of
(\ref{EOMm}) with $M^2_{V,0}=0$. This solution
corresponds to 4D effective theory gauge fields.  It has a finite
kinetic energy, and trivially satisfies (\ref{HC2}).  The fact that such gauge
fields have  a constant
transverse profile guarantees charge universality of fermions in
the 4D effective theory (see below).

To find the massive mode solutions, we can express (\ref{EOMm}) in
terms of $u$ and $\psi$ and obtain a Schroedinger equation:
\be \left(-\partial^2_u+V\right)\psi=M^2_V\psi,
\label{SchroedingerV}\ee
where the ``potential'' is
\be
V(u)=e^{A-B}\left(m+e_V\mathcal{A}_{\varphi}\right)^2+e^{-(2\phi+A+B)/4}
\partial^2_u
e^{(2\phi+A+B)/4}. \ee
We want to find the complete set of solutions to
(\ref{SchroedingerV}) satisfying the NC (\ref{NCV}) and the
boundary conditions (\ref{HC2}), which can be written in terms of
$u$ and $\psi$ as follows
\be \left(\lim_{u\rightarrow \overline{u}}-\lim_{u\rightarrow
0}\right)
\left\{\psi^*\left[-\partial_u+\frac{1}{4}\left(2\partial_u\phi +
\partial_uA+\partial_u
B\right)\right]\psi\right\}=0. \label{HCV1}\ee
In order for (\ref{HCV1}) to be satisfied, both the limits
$u\rightarrow 0$ and $u\rightarrow \overline{u}$ must be finite.
Condition (\ref{HCV1}) ensures that the Hamiltonian in the
Schroedinger equation (\ref{SchroedingerV}) is hermitian, and so has real
eigenvalues and an orthonormal set of eigenfunctions.  Therefore, we
shall call it the hermiticity condition (HC).

So far our analysis has been valid for all axially symmetric
solutions of the form (\ref{axisymmetric}).  We will now use these
results to determine the fluctuation spectrum about the
conical-GGP solution (\ref{GGPsolution}). We observe that $V(u)$
then contains a delta-function contribution, arising from the
second-order derivative of the conical metric function
$\partial_u^2B$ (see Appendix \ref{deltafn} and eq. (\ref{B2u})). However we can
drop it because $\partial_u^2B$ also contains stronger singularities
at $u=0$ and $u=\overline{u}$: respectively $1/u^2$ and
$1/(\overline{u}-u)^2$. These singularities are also a consequence
of the behaviour of $e^B$ given in (\ref{assumeeB}) and they imply
that the behaviour of the wave functions close to $u=0$ and
$u=\overline{u}$ cannot depend on the mass. In particular, this
immediately implies that if the wave functions of zero modes are peaked near to
one of the branes, then the same will be true also for the infinite tower of
non-zero modes.  In
other words, we cannot hope to dynamically generate a brane world
scenario, in which zero modes are peaked on the brane, and
massive modes are not, leading to weak coupling between the two
sectors\footnote{In fact, a similar singular behaviour for the potential in
general arises for the general axisymmetric solutions
given in \cite{GGP} and studied in \cite{Burgess:2004dh}, where the hypothesis
(\ref{assumeeB}) is relaxed.}. If we are to interpret the zero mode gauge fields
as those of the SM, therefore, for the massive modes to have escaped
detection they must have a large mass gap.

Meanwhile, we note that in contrast to the non-relativistic quantum mechanics
problem, here we cannot deduce qualitative results about the mass spectrum from
the shape of the potential.  This is because the boundary conditions to be
applied in the context of dimensional reduction are in general different to
those in problems of quantum mechanics.  In particular, the HC (\ref{HCV1}) is a
non-linear condition, contrary to the less general linear boundary conditions
usually encountered in quantum mechanics to ensure hermiticity of the
Hamiltonian.  We will be able to impose the more general case thanks to the
universal asymptotic behaviour of the Kaluza-Klein tower.

Returning then to our explicit calculation of the Kaluza-Klein
spectrum, we can write $V(u)$ as
\be V(u)=V_0+v\cot^2\left(\frac{u}{r_0}\right)
+\overline{v}\tan^2\left(\frac{u}{r_0}\right),\label{SchroedingerVGGP}\ee
and
\be r_0^2V_0 \equiv 2m\omega (m-N_V)\ob -\frac{3}{2},\quad
r_0^2v\equiv m^2 \omega^2-\frac{1}{4},\quad r_0^2 \overline{v}\equiv
(m-N_V)^2 \ob^2-\frac{1}{4}.\ee
Moreover in this case the expression (\ref{HCV1}) for the HC becomes
\be \lim_{u\rightarrow
\overline{u}}\psi^*\left(-\partial_u+\frac{1}{2}\frac{1}{u-\overline{u}}
\right)\psi
-\lim_{u\rightarrow
0}\psi^*\left(-\partial_u+\frac{1}{2u}\right)\psi=0.
\label{HCV2}\ee
If we introduce $z$ and $y$ in the following way \cite{manning}
\be z=\cos^2\left(\frac{u}{r_0}\right),\qquad
\psi=z^{\gamma}\left(1-z\right)^{\beta}y(z),\ee
equation (\ref{SchroedingerV}) becomes
\be z(1-z)\partial_z^2y+\left[c-(a+b+1)z\right]\partial_zy-ab
y=0\label{hyper},\ee
where
\bea \gamma &\equiv& \frac{1}{4}\left[1+2(m-N_V)\ob\right], \,\,
\beta\equiv \frac{1}{4}\left(1+2m\omega\right),
\,\, c\equiv 1+(m-N_V)\ob, \nonumber \\
a&\equiv&\frac{1}{2}+\frac{m}{2}\omega+\frac{1}{2}(m-N_V)\ob
+\frac{1}{2}\sqrt{r_0^2M^2_{V,m}+1+\left[m\omega-(m-N_V)\ob\right]^2},\nonumber\\
b &\equiv&\frac{1}{2}+\frac{m}{2}\omega+\frac{1}{2}(m-N_V)\ob
-\frac{1}{2}\sqrt{r_0^2M^2_{V,m}+1+\left[m\omega-(m-N_V)\ob\right]^2},
\label{gbetaabcV}\eea
and
\be \omega\equiv(1-\delta/2\pi)^{-1}, \qquad
\ob\equiv(1-\overline{\delta}/2\pi)^{-1}.\ee
Equation (\ref{hyper}) is the hypergeometric equation and its
solutions are known. For $c\neq 1$ the general solution is a
linear combination of the following functions:
\be y_1(z)\equiv F(a,b,c,z), \quad y_2(z)\equiv z^{1-c}
F(a+1-c,b+1-c, 2-c, z), \label{y12}\ee
where $F$ is Gauss's hypergeometric function.  So for $c\neq 1$
the general integral of the Schroedinger equation is
\be\psi=K_1\psi_1 + K_2 \psi_2,\label{cnot1}\ee
where
\be \psi_i\equiv z^{\gamma}(1-z)^{\beta}y_i. \label{psii} \ee
and $K_{1,2}$ are integration constants. For $c=1$ we have
$\psi_1=\psi_2$ but we can construct a linearly independent solution
using the Wronskian method and the general solution reads
\be\psi=K_1\psi_1 + K_2
\psi_1\int^u\frac{du'}{\psi_1^2(u')}.\label{wronskian}\ee

Now we must impose the NC (\ref{NCV}) and
HC (\ref{HCV2}), to select the physical modes.  In Appendix
\ref{boundcond} we give explicit calculations; the final result is
that the NC and HC give the following discrete spectrum. The wave
functions are
\bea \psi&\propto&  z^{\gamma}(1-z)^{\beta}F(a,b,c,z), \quad for\quad m\geq
N_V,\label{psi1}\\
\psi&\propto&  z^{\gamma +1-c}(1-z)^{\beta} F(a+1-c,b+1-c, 2-c,
z), \quad for\quad m<N_V.\label{psi2} \eea
and the squared masses are as follows:
\begin{itemize}
\item For  $N_V\leq m < 0$
\be
M_{V\,n,m}^2=\frac{4}{r_0^2}\left\{n(n+1)+\left(\frac{1}{2}+n\right)\left[
-m\omega+(m-N_V)\ob\right]\right\}>0.\label{MV1}\ee
\item For  $m\geq N_V$ and $m\geq 0$
\be
M_{V\,n,m}^2=\frac{4}{r_0^2}\left\{n(n+1)+\left(\frac{1}{2}+n\right)\left[
m\omega+(m-N_V)\ob\right]+m\omega(m-N_V)\ob\right\}\geq
0.\label{MV2}\ee
\item For $m<N_V$ and $m<0$
\be
M_{V\,n,m}^2=\frac{4}{r_0^2}\left\{n(n+1)+\left(\frac{1}{2}+n\right)\left[
-m\omega+(N_V-m)\ob\right]-m\omega(N_V-m)\ob\right\}>
0.\label{MV3}\ee
\item For  $0\leq m <N_V$
\be
M_{V\,n,m}^2=\frac{4}{r_0^2}\left\{n(n+1)+\left(\frac{1}{2}+n\right)\left[
m\omega+(N_V-m)\ob\right]\right\}>0.\label{MV4}\ee
\end{itemize}
The masses given in (\ref{MV1}) and (\ref{MV2}) correspond to the
wave function (\ref{psi1}) whereas the masses given in (\ref{MV3})
and (\ref{MV4}) correspond to the wave function (\ref{psi2}).
We observe that there are no tachyons and that the only zero mode is for $n=0$,
$m=0$
and $N_V=0$ ($e_V=0$), corresponding to gauge fields in the 4D low energy
effective theory.

As a check, we can consider the $S^2$ limit ($\omega,\ob \rightarrow
1$), whose mass spectrum is well-known.  Our spectrum
(\ref{MV1})-(\ref{MV4}) reduces to
\be a^2M^2_V=l(l+1)-\left(\frac{N_V}{2}\right)^2,\quad
multiplicity=2l+1 \, ,\ee
where\footnote{The number $l$ is defined in different ways in
equations (\ref{MV1})-(\ref{MV4}). For instance we have $l\equiv
n+|N_V/2|$ for (\ref{MV1}).} $l=|\frac{N_V}{2}|+k$ and
$k=0,1,2,3,...$. This is exactly the result that one finds by
using the spherical harmonic expansion \cite{sphere} from the beginning.

At this stage we can point towards a novel property of the final mass
spectrum.  Observe that in the large $\a$ (small $\ob$) limit the volume $V_2$
given in equation (\ref{V2})
becomes large but the mass gap between two consecutive Kaluza-Klein
states does not reduce to zero as in standard Kaluza-Klein
theories\footnote{This is also true for the proper volume of the 2D internal
manifold.}. This a consequence of the shape of our background
manifold and in particular of the conical defects.  Notice that the large $\a$
limit corresponds to a negative tension brane at $u=\bar{u}$, but not
necessarily at $u=0$.

In
Section \ref{fermion} we will show that the same effect appears
also in the fermionic sector, and we will turn to a discussion of
its implications in Section \ref{LargeMV}.

\subsection{4D Effective Gauge Coupling}
Let us end the discussion on gauge fields by briefly presenting the 4D effective
gauge coupling.  This can be obtained by dimensionally reducing the 6D gauge
kinetic term.  We consider the zero mode fluctuations in ${\cal H}$, about the
background (\ref{axisymmetric}), so that the 4D effective gauge kinetic term is:
\bea
&&\int d^6X \sqrt{-G}\left\{-\frac{1}{4{g}^2} e^{\kappa \sigma/2}
TrF_{MN}F^{MN}\right\} \rightarrow \nonumber \\
&& \phantom{0000000000000000} \int d^4x\left\{ -\frac{1}{4{g}^2}\left[\int
dud\varphi e^{(3A+B)/2} f^{2}_{0} \right]TrF_{\mu\nu}F^{\mu\nu} \right\} \, .
\eea
Recalling that $f_0 = const$ and normalizing it to one, we can read:
\be
\frac{1}{g_{eff}^2} = \frac{1}{g^2} V_2 \, .
\ee

\section{Fermions}\label{fermion}
\setcounter{equation}{0}

We will now consider fermionic perturbations, and in particular
our interest will be in the sector charged under the 4D effective
gauge group, ${\cal H}$, discussed above.  These fields arise from
the hyperinos and the ${\cal H}$ gauginos, for which we also
restrict ourselves to those orthogonal to the $U(1)_R$.  Thus we
are considering matter charged under the non-Abelian gauge
symmetries of the 4D effective theory. For instance, for the
anomaly free model $E_6 \times E_7 \times U(1)_R$, with the
monopole embedded in the $E_6$, the gauginos in the {\bf $78$} of
$E_6$ contain a {\bf $16$} + {\bf ${\overline{16}}$} fundamental representation
of
the grand unified gauge group $SO(10)$, and our analysis will be applicable to
them.  In Tabel \ref{T:summary}, we give the complete list of fermion fields
that are included in our study, for the said example.

 We proceed in much
the same way as for the gauge field sector of the previous section,
transforming the dynamical equations and necessary boundary
conditions into a Schroedinger-like problem, to obtain the physical
modes and discrete mass spectrum.

The bilinear action for the fluctuations of interest takes a particularly
simple form, comprising as it does of the standard Dirac action:
\be S_F=\int d^6X \sqrt{-G}\,\,\overline{\lambda}\Gamma^M
D_M\lambda, \label{1/2action}\ee
where\footnote{Our conventions for $\Gamma^A$ and $
\Omega^{[A,B]}_M$ are given in Appendix \ref{conventions}.}
\be
D_M\lambda=\left(\partial_M+\frac{1}{8}\Omega^{[A,B]}_M[\Gamma_A,\Gamma_B]
+ie\mathcal{A}_M\right)\lambda.
\label{Dlambda} \ee
Here $e$ is the charge of $\lambda$ under the $U(1)$ monopole, and $G_{MN}$,
$\Omega^{[A,B]}_M$ and $\mathcal{A}_M$ are the background metric, spin
connection and gauge field corresponding to an axisymmetric
solution (\ref{axisymmetric}).  Analogously to the gauge field
analysis, in order to derive the Dirac equation
\be \Gamma^MD_M \lambda = 0 \label{DiracEq}\ee
from (\ref{1/2action}) by using an action principle
\cite{Wetterich:1984dz}, we require conservation of fermionic
current:
\be \int d^6X
\partial_M\left(\sqrt{-G}\,\,\overline{\lambda}\Gamma^M\lambda\right)=0.
\label{hermiticity}\ee
Equation (\ref{hermiticity}) implies that the Dirac operator
$\Gamma^M D_M$ is hermitian, and we shall again refer to it
as the HC.  Our aim is to find the complete fermionic spectrum, that
is a complete set of normalizable solutions of (\ref{DiracEq})
satisfying (\ref{hermiticity}).

Some care is now needed when discussing the background felt by the
fermionic sector in (\ref{Dlambda}).  As already mentioned, in order to have a
correctly defined gauge connection, it is necessary to use two patches related
by a single-valued gauge transformation.  The same is true for the spin
connection, which must be defined in such a way as to imply the conical defects
in the geometry.
Henceforth we focus on the patch including the $\rho=0$ brane,
chosen to be $0 \leq \rho < {\bar \rho}$.  For this patch a good
choice for the vielbein is
\be e^{\alpha}_{\mu}=e^{A/2}\delta^{\alpha}_{\mu}, \quad
\{e^a_m\}=\left(\ba {cc} \cos\varphi & -e^{B/2}\sin\varphi
\\ \sin\varphi & e^{B/2}\cos\varphi \ea \right), \label{vielbein}\ee
where $\alpha$ is a 4D flat index, $a=5,6$ a 2D flat index and
$m=\rho, \varphi$. The corresponding spin connection is
\bea
\Omega_{\mu}^{[\alpha,5]}&=&\frac{1}{2}A'e^{A/2}\delta^{\alpha}_{\mu}\cos\varphi
,
\quad
\Omega_{\mu}^{[\alpha,6]}=\frac{1}{2}A'e^{A/2}\delta^{\alpha}_{\mu}\sin\varphi,
\nonumber\\
\Omega_{\rho}^{[5,6]}&=&0,\qquad
\Omega\equiv\Omega_{\varphi}^{[5,6]}=\left(1-\frac{1}{2}B'e^{B/2}\right),
\label{spinconnection} \eea
where $'\equiv \partial_{\rho}$.  It can be checked that this gauge choice
correctly reproduces Stokes' theorem for a small domain including the conical
defect\footnote{See Appendix \ref{deltafn} for some steps in this
calculation.}.

We are now ready to study the Dirac equation (\ref{DiracEq}) for
6D fluctuations, and write it in terms of 4D effective fields.
 Since $\lambda$ is a 6D Weyl spinor we can represent it by
\be \lambda=\left(\ba {c} \lambda_4 \\ 0 \ea\right), \ee
where $\lambda_4$ is a 4D Dirac spinor:
$\lambda_4=\lambda_R+\lambda_L$, $\gamma^5\lambda_R=\lambda_R,$
$\gamma^5\lambda_L=-\lambda_L$. By using the ansatz
(\ref{axisymmetric}), the vielbein (\ref{vielbein}), the spin
connection (\ref{spinconnection}) and our conventions for
$\Gamma^A$ in Appendix \ref{conventions}, the Dirac equation
(\ref{DiracEq}) becomes
\bea e^{-A/2}\gamma^{\mu}\partial_{\mu}\lambda_L&=&
e^{i\varphi}\left[-\partial_{\rho}-ie^{-B/2}\left(\partial_{\varphi}+ie\mathcal{
A}_{\varphi}\right)
-A'+\frac{1}{2}\Omega e^{-B/2}\right]\lambda_R, \label{EqR}\\
 e^{-A/2}\gamma^{\mu}\partial_{\mu}\lambda_R&=&
e^{-i\varphi}\left[\partial_{\rho}-ie^{-B/2}\left(\partial_{\varphi}+ie\mathcal{
A}_{\varphi}\right)
+A'-\frac{1}{2}\Omega e^{-B/2}\right]\lambda_L.\label{EqL}\eea
Performing the Fourier mode decomposition:
\be
\lambda_4(X)=\lambda_R(X)+\lambda_L(X)=\sum_m\left(\lambda_{R,m}(x)f_{R,m}(\rho)
+\lambda_{L,m}(x)f_{L,m}(\rho)\right)e^{im\varphi},\label{Fourier}\ee
where $m$ is an integer, and inserting into (\ref{EqR}) and
(\ref{EqL}) we find:
\bea e^{-A/2}\gamma^{\mu}\partial_{\mu}\lambda_{L,m+1}f_{L,m+1}&=&
\left[-\partial_{\rho}+e^{-B/2}\left(m+\frac{1}{2}\Omega
+e\mathcal{A}_{\varphi}\right)
-A'\right]\lambda_{R,m}f_{R,m}, \label{EqRm}\\
 e^{-A/2}\gamma^{\mu}\partial_{\mu}\lambda_{R,m-1}f_{R,m-1}&=&
\left[\partial_{\rho}+e^{-B/2}\left(m-\frac{1}{2}\Omega
+e\mathcal{A}_{\varphi}\right)
+A'\right]\lambda_{L,m}f_{L,m}.\label{EqLm}\eea

For the boundary conditions, analogously to the gauge fields, the NC can be
found to be:
\be \int du \left|\psi\right|^2 <\infty \label{NC}\ee
where
\be \psi\equiv e^{A+B/4}f_{Rm} \label{psi}\ee
and a similar condition for left-handed spinors.
Meanwhile, the HC (\ref{hermiticity}) can be written:
 \be
\left[\sqrt{-G}\,\,f_{L,m+1}f_{R,m}^{*}
\right]_0^{\overline{\rho}}=0.\label{HC3}\ee

Having set up the dynamical equations and the relevant boundary
conditions, we shall now use this information to study the
complete fermionic spectrum in Subsections \ref{zero} and
\ref{complete}. In particular, we will focus on the questions of
wave function localization, and the mass gap problem, crucial to
the development of a phenomenological brane world model.

\subsection{Zero Modes}\label{zero}
We begin by finding the zero mode solutions, for which the problem
simplifies considerably.  Indeed, for the zero modes
$\gamma^{\mu}\partial_{\mu}=0$, and the equations for right- and
left-handed modes (\ref{EqRm}) and (\ref{EqLm}) decouple:
\bea
\left[\partial_{\rho}-e^{-B/2}(m+e\mathcal{A}_{\varphi})+A'-\frac{1}{2}\Omega
e^{-B/2}\right]f_{R,m}&=&0,\label{EqRzero}\\
\left[\partial_{\rho}+e^{-B/2}(m+e\mathcal{A}_{\varphi})+A'-\frac{1}{2}\Omega
e^{-B/2}\right]f_{L,m}&=&0.\label{EqLzero}\eea
By using the expression for $\Omega$ in equation
(\ref{spinconnection}), the solution of (\ref{EqRzero}) is
\be f_{R,m}(\rho)\propto\exp\left[-A-\frac{1}{4}B
+\int^{\rho}d\rho'
e^{-B/2}\left(m+\frac{1}{2}+e\mathcal{A}_{\varphi}\right)\right],
\label{zerosolution}
\ee
whereas the solution of (\ref{EqLzero}) can be obtained by
replacing $m,e\rightarrow -m,-e$ in (\ref{zerosolution}). The
solution (\ref{zerosolution}) for $e=0$ was found in
\cite{Schwindt:2003er}.  Here we give the expression for every $e$
because we want to include charged fermions. We note that the zero
mode solution (\ref{zerosolution}) automatically satisfies the HC
given in (\ref{HC3}). From (\ref{zerosolution}), (\ref{psi}) and
(\ref{u}) we obtain
\be \psi \propto\exp\left[\int^{u}du'
e^{(A-B)/2}\left(m+\frac{1}{2}+e\mathcal{A}_{\varphi}\right)\right].
\label{psizerosolution} \ee
For the conical-GGP background, (\ref{GGPsolution}), the explicit
expression for $\psi$ is
\be \psi\propto
\sin\left(\frac{u}{r_0}\right)^{\omega(1/2+m)}\cos\left(\frac{u}{r_0}\right)^{
\ob(N-1/2-m)},
\label{0GGP}\ee
where we used (\ref{DiracQ}).
 The NC is satisfied when
\be \frac{\delta}{4\pi}-1<m<N-\frac{\overline{\delta}}{4\pi}.
\label{NCGGP}\ee
From here we retrieve the  result that for the sphere, which has
$\delta=\overline{\delta}=0$, there exist normalizable zero modes
only for $e\neq 0$ ($N\neq0$), that is for a non-vanishing monopole
background \cite{sphere}.  Moreover, as found in
\cite{Schwindt:2003er}, we see that the conical defects also make
massless modes possible, provided that there is at least one
negative deficit angle, even if $N=0$.  However, (\ref{NCGGP})
implies that for positive tension branes, $\delta, {\overline
\delta} >0$, the adjoint of ${\cal H}$, which has $e=0$, is
projected out.  If ${\cal H}$ contains the SM gauge group, this is
appealing since the fermions of the SM are not in adjoint
representations.  In any case, the number of families depends on
$\delta$, ${\overline \delta}$ and $N$.

Let us now consider the wave function profiles, (\ref{0GGP}).  Observe that
$\psi$ is peaked on the $u=0$ brane (that is, $\psi \rightarrow
\infty$ as $u \rightarrow 0$, and $\psi \rightarrow 0$ as $u \rightarrow
\bar{u}$) when
\be m<-1/2, \qquad and \qquad m<-1/2+N. \label{LCGGP}\ee
By comparing (\ref{NCGGP}) and (\ref{LCGGP}) we understand that we
have normalizable and peaked $\psi$ only for $\delta<0$ (that
is for negative tension brane). If $N-\overline{\delta}/4\pi>0$ we
can have normalizable zero modes for $\delta>0$ (positive tension
brane) but the corresponding $\psi$ are not peaked on the $u=0$
brane (indeed, $\psi \rightarrow 0$ as $u \rightarrow 0$). On the
other hand from (\ref{0GGP}) we see that $\psi$ is peaked on
the $u=\overline{u}$ brane when
\be m>-1/2, \qquad and \qquad m>-1/2+N. \label{LCGGPbar}\ee
By comparing (\ref{NCGGP}) and (\ref{LCGGPbar}) we understand that
we have normalizable and peaked $\psi$ only for
$\overline{\delta}<0$ (that is for negative tension brane).

We can also analyze the chirality structure. Since the left handed
wave functions can be obtained from (\ref{0GGP}) by replacing
$m,N\rightarrow -m,-N$, in order for them to be peaked on the
$u=0$ brane we need
\be m>1/2, \qquad and \qquad m>1/2+N, \ee
whereas in order for the left handed wave functions to be
peaked on the $u=\overline{u}$ brane we need
\be m<1/2, \qquad and \qquad m<1/2+N. \ee
So, if we were to require that $\psi$ be peaked on a brane, we
always have a chiral massless spectrum because the chirality index
counts the difference of modes in $f_{R,m}$ and $f_{L,m}$ with
given $m$, $N_R(m)-N_L(m)$ \cite{Schwindt:2003er}. We should point
out, however, that in fact peaked zero modes, $\psi$, may not be
necessary in order to have an acceptable phenomenology. The answer
to this question can be found only after constructing the complete
spectrum, and studying the couplings between different 4D
effective fields.

\subsection{Massive Modes}\label{complete}
We now move on to a study of the complete Kaluza-Klein tower for
the fermions.  We begin by establishing the corresponding
Schroedinger problem.  The two coupled first order ODEs, equations
(\ref{EqRm}) and (\ref{EqLm}), can be equivalently expressed as a
single second order ODE and a constraint equation, as
follows\footnote{Eq.(\ref{2nd}) can also be obtained by squaring the 6D Dirac
operator, and using that $[D_M,D_N] = \frac{1}{4}
R_{MN}^{\,\,\,\,\,\,\,\,\,\,AB}\Gamma_{AB} + ie F_{MN}$.}
\bea
&&e^A\left(-\partial_{\rho}^2+h\partial_{\rho}+g_m\right)f_{R,m}=M_{F,m}^2f_{R,m
},\label{2nd}
\\
&&M_{F,m}f_{L,m+1}=e^{A/2}\left[-\partial_{\rho}-A'+\left(m+\frac{1}{2}\Omega
+e\mathcal{A}_{\varphi}\right)e^{-B/2}\right]f_{R,m}, \label{fLR}
\eea
where $M_{F,m}^2$ are the eigenvalues of
$\left(\gamma^{\mu}\partial_{\mu}\right)^2$ and
\bea h &\equiv&-\frac{5}{2}A'+(\Omega-1)e^{-B/2},\\
g_m &\equiv&\left[\frac{1}{2}\Omega'-\frac{1}{4}\Omega
B'-\frac{m}{2}B'
+\frac{5}{4}A'\Omega+\left(\frac{m}{2}-1\right)A'-\frac{e}{2}B'\mathcal{A}_{
\varphi}+
e\mathcal{A}_{\varphi}'+\frac{e}{2} A'
\mathcal{A}_{\varphi}\right]e^{-B/2}\nonumber\\
&&+\left[m(m+1)+\frac{1}{2}\Omega-\frac{\Omega^2}{4}
+(2m+1)e\mathcal{A}_{\varphi}+e^2\mathcal{A}_{\varphi}^2\right]e^{-B}-A''-\frac{
3}{2}(A')^2.
\label{g}\eea
Once $f_{R,m}$ is known we can compute $f_{L,m+1}$ by using
(\ref{fLR}), so we can focus on $f_R$ and study the second order
ODE (\ref{2nd}). If we express this equation in terms of $\psi$
and $u$ we obtain the Schroedinger equation
\be
\left(-\partial^2_u+V\right)\psi=M^2_{F,m}\psi,\label{Schroedinger1}\ee
where the ``potential'' $V$ is given by
\bea
V(u)&=&e\partial_u\mathcal{A}_{\varphi}e^{(A-B)/2}+\left(\frac{1}{2}+m
+e\mathcal{A}_{\varphi}\right)\partial_u e^{(A-B)/2}\nonumber \\
&&+\left[\frac{1}{4}+m+e\mathcal{A}_{\varphi}+\left(m+e\mathcal{A}_{\varphi}
\right)^2\right]e^{A-B}.\eea
We observe that transformation (\ref{psi}) exactly removes the delta-functions
which appear in (\ref{g}) through $\Omega'$ (see Appendix \ref{deltafn}).
However, just as for the gauge fields, a singular behaviour is observed in the
potential, so that the asymptotic behaviour of the wave functions does not
depend on their mass.

Our problem
is now reduced to solving equation (\ref{Schroedinger1}) with
the conditions NC (\ref{NC}) and HC (\ref{HC3}). By using
(\ref{fLR}) for $M_F\neq 0$
 and
definitions (\ref{u}) and (\ref{psi}) we can rewrite (\ref{HC3})
as follows
\be \left(\lim_{u\rightarrow \overline{u}}-\lim_{u\rightarrow
0}\right)
\psi^*\left[-\partial_u+\left(m+\frac{1}{2}+e\mathcal{A}_{\varphi}\right)e^{
(A-B)/2}\right]\psi=0.\label{HC5}\ee

We can now proceed in exactly the same way as for the gauge field
sector.  For the conical-GGP solution (\ref{GGPsolution}) the
explicit expression for $V$ has the form (\ref{SchroedingerVGGP}),
but now with
\bea r_0^2V_0&\equiv& \left(\frac{1}{2}+m\right)\left[\ob -\omega+2\omega \ob
\left(\frac{1}{2}+m-N\right)\right]-\ob N,\\
r_0^2v&\equiv&
\left(\frac{1}{2}+m\right)\left[-\omega+\omega^2\left(\frac{1}{2}+m\right)\right],\\
r_0^2
\overline{v}&\equiv&\left(\frac{1}{2}+m\right)\left[\ob+\ob^2
\left(\frac{1}{2}+m-2N \right)\right] +\ob N\left(\ob
N-1\right). \eea
Moreover, in this case the explicit expression for the HC is
\be \lim_{u\rightarrow
\overline{u}}\psi^*\left(-\partial_u+\ob\frac{m+1/2-N}{\overline{u}-u}
\right)\psi
-\lim_{u\rightarrow
0}\psi^*\left(-\partial_u+\omega\frac{m+1/2}{u}\right)\psi=0.
\label{HC4}\ee
As in the gauge fields sector we introduce $z$ and $y$ in the
following way
\be z=\cos^2\left(\frac{u}{r_0}\right),\qquad
\psi=z^{\gamma}\left(1-z\right)^{\beta}y(z), \label{psiz} \ee
so that equation (\ref{SchroedingerVGGP}) becomes a hypergeometric
equation (\ref{hyper}), with parameters:
\bea \gamma &\equiv&
\frac{1}{2}\left[1+\ob\left(\frac{1}{2}+m-N\right)\right], \,\,
\beta\equiv
\frac{1}{2}\left[1-\omega\left(\frac{1}{2}+m\right)\right],
\,\, c\equiv \frac{3}{2}+\ob \left(\frac{1}{2}+m-N\right), \nonumber \\
a&\equiv&1+\frac{\ob}{2}
\left(\frac{1}{2}+m-N\right)-\frac{\omega}{2}\left(\frac{1}{2}+m\right)+\frac{1}{2}
\sqrt{\Delta},\nonumber\\
b &\equiv&1+\frac{\ob}{2}
\left(\frac{1}{2}+m-N\right)-\frac{\omega}{2}\left(\frac{1}{2}+m\right)-\frac{1}{2}
\sqrt{\Delta},\nonumber\\
\Delta&\equiv& r_0^2 M^2_{F,m}+\left(\ob N\right)^2+
\left(\frac{1}{2}+m\right)
\left[\ob(\ob-2\omega)\left(\frac{1}{2}+m-N\right)-\ob^2N+\omega^2\left(\frac{1}{2}
+m\right)\right].\nonumber\eea
We can construct two independent solutions $\psi_1$ and $\psi_2$
of the Schroedinger equation (\ref{SchroedingerVGGP}) as in Section
\ref{gaugefields}, and impose the NC (\ref{NC}) and the HC
(\ref{HC4}) to obtain the physical modes.  The resulting wave
functions are:
\be\psi=K_1\psi_1 + K_2 \psi_2,\ee
where the integration constants, $K_{1,2}$, are fixed in Appendix
\ref{fermspectrum}.
We plot a few of the wave function profiles in Figure
\ref{fig:wavefns}.

The complete discrete mass spectrum is also given
in Appendix \ref{fermspectrum}.  There it can be seen that the same finiteness
of the mass gap in the large $\alpha$ (hence large volume) limit, found in the
gauge field spectrum, can be observed here.   Moreover, for $\alpha \sim 1$, the
mass gap between the zero modes and the massive states now goes as:
\be M^2_{GAP}\sim \frac{1}{r_0^2} + \frac{1}{r_1^2}. \ee
Therefore, for the fermions, a finite mass gap in the large volume limit can
also be obtained by taking $r_0 \rightarrow \infty$ and turning on  $\delta$,
thus allowing $r_1$ to remain finite.
This contrasting behaviour to the standard Kaluza-Klein picture is a
consequence of the conical defects ($\ob \neq 1$ and $\omega\neq 1$) in
our internal manifold.  Below we shall consider its implications
for phenomenology.

\begin{figure}
\centering
\begin{tabular}{cc}
\epsfig{file=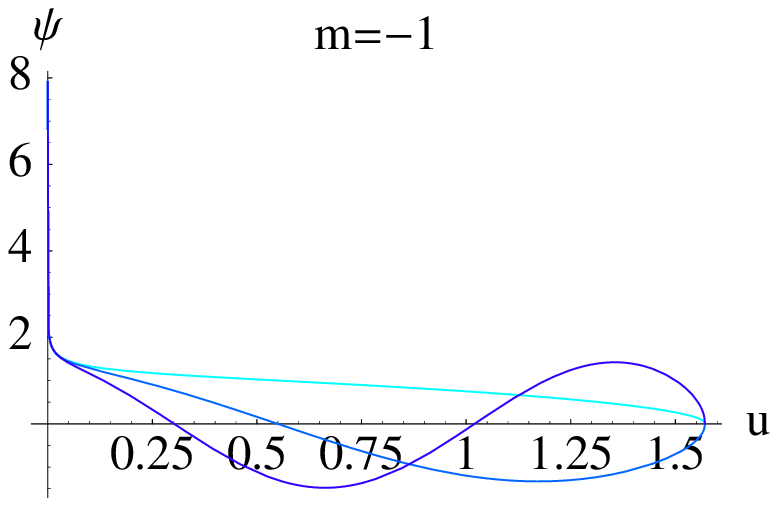,width=0.45\linewidth,clip=} &
\epsfig{file=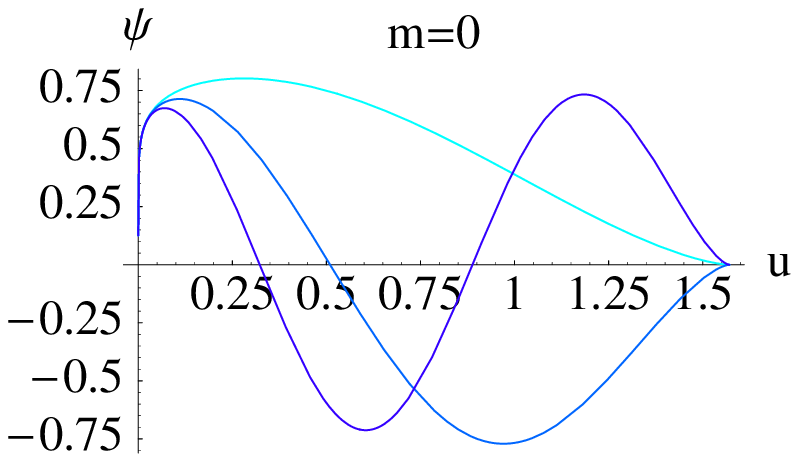,width=0.45\linewidth,clip=}
\end{tabular}
\caption{\footnotesize Fermion Wave Function Profiles: $n=0,1,2$
modes plotted for angular momentum numbers $m=-1,0$ (eqs
(\ref{02}) and (\ref{m2}) respectively). The parameters are chosen
to be $(r_0, \omega, \overline{\omega},e)=(1,1/4,1,0)$,
corresponding to a single negative tension brane at $u=0$. Also the
normalisation constant is set to 1. The number of intersections
with the $u$-axis equals $n$, according to quantum mechanics. Notice
that the $(m,n)=(-1,0)$ mode is massless, and that given a
localized massless mode, there is also an infinite Kaluza-Klein
tower of localized massive modes.}\label{fig:wavefns}
\end{figure}

\subsection{4D Effective Fermion Charges}
Let us first end this section on fermion fluctuations by obtaining
their 4D effective gauge couplings.  This can be calculated by
going beyond their bilinear Lagrangian, and considering the interaction term:
\be  \int d^6X \sqrt{-G}\,\,\overline{\lambda}\Gamma^M
D_M\lambda = \dots + \int d^6X \sqrt{-G}\,\,\overline{\lambda}\Gamma^{\mu}
(\partial_{\mu} + e V_{\mu}) \lambda + \dots \, .
\ee
Using the results for the Kaluza-Klein decomposition found in the preceding
sections
\bea \lambda(X) &=& \sum_{m,n} \lambda_{mn}(x) f^{(\lambda)}_{mn}(\rho)
e^{im\varphi} = \sum_{m,n} \lambda_{mn}(x) \psi^{(\lambda)}_{mn}(u) e^{-A -B/4}
e^{im\varphi} \, ,\nonumber \\
V_{\mu}(X) &=& \sum_{m,n} V_{\mu \, mn}(x) f^{(V)}_{mn}(\rho) e^{im\varphi} =
\sum_{m,n} V_{\mu \, mn}(x) \psi^{(V)}_{mn}(u) e^{-(3A+B)/4} e^{im\varphi} \, ,
\eea
and recalling that the gauge field zero mode is $f^{(V)}_{00} = 1$, a fermion
mode $\lambda_{mn}$ has the following coupling to the 4D effective gauge group:
\bea
e_{eff} &=& \frac{e \, \int d\varphi du \, \overline{\psi^{(\lambda)}_{mn}} \,
f^{(V)}_{00} \, \psi^{(\lambda)}_{mn}}{\int d\varphi du \,
\overline{\psi^{(\lambda)}_{mn}}\, \psi^{(\lambda)}_{mn}} \nonumber \\
&=& e \, .
\eea
Since the gauge field zero mode has a constant wave profile, the effective
charges for the fermion modes are universal.  This is a general result, and
independent of any possible localization properties of the fermion modes:
massless and massive fermion modes will always have the same coupling to the
massless gauge fields.  Again, we find that a large mass gap is required in the
fermion spectrum in order to hide the Kaluza-Klein tower.  We now consider this
issue in more detail.

\section{Large Volume Compactifications with a Large Mass Gap}\label{LargeMV}

In the previous two sections, we have calculated the complete
Kaluza-Klein spectrum for the warped brane world compactification
of 6D supergravity, for two interesting sectors of the gauge and
matter fluctuations. We are now ready to consider the possible
implications of our results.

6D brane world models have long been of interest in the context of
Large Extra Dimensions, since these may help with the gauge
hierarchy problem.  In the conventional ADD picture \cite{ADD}, Standard
Model particles must be confined to a 4D brane world, in order to
explain why the large extra dimensions have escaped detection.  It
would certainly be of interest to develop a dynamical description
of this localization, within the context of low energy effective
field theory.

This could be achieved, for instance, if the zero mode wave profiles
were peaked near to a brane, and the heavy modes suppressed there
\cite{Randjbar-Daemi:2003qd}. However, we have found that zero mode
fermions can be peaked near to negative tension branes, only at the
price of localizing the whole Kaluza-Klein tower (see Figure
\ref{fig:wavefns}). Therefore, strong couplings are expected between
light and heavy modes.  If the zero mode bulk fermions are to be
interpreted as matter in the SM, then apparently the only way to
explain why we do not observe all the Kaluza-Klein modes is by
insisting that their mass gap is larger than the $100GeV$ scale
probed to date.

Usually, this would bring us back to the classical Kaluza-Klein
scenario, with the extra dimensions required to be very small  (at
least $(100GeV)^{-1}$ scale), and the generation of the gauge
hierarchy lost. However, in our framework we have seen that a
large mass gap can occur even if the volume $V_2$, defined as the
ratio $\kappa^2/\kappa_4^2$, becomes large. In the fermionic
sector this set up can be achieved with the conical defect associated
with the warping ($\omega\neq 1$ and $\overline{\omega}=1$), by
taking the parameter $r_1$ to be small.
 Another way is turning on the other defect ($\ob\neq
1$) and then taking the large $\alpha$ limit, that is small $\ob$
limit, corresponding to a negative tension brane.  The volume $V_2$ in
(\ref{V2}) becomes large but, for both
the fermions and gauge fields, the mass gap does not reduce to
zero. We observe that the latter mechanism works also for
$r_0=r_1$, that is $\omega=\overline{\omega}$, which corresponds
to the unwarped ``rugby ball'' solution with branes of equal tension at each
of the two poles \cite{carroll, sled}.

The general idea of relaxing the phenomenological constraints on
the size of the extra dimensions by deforming the shape of the
internal space was proposed in \cite{Dienes:2001wu}.  There, it
was shown that the presence of shape moduli can imply that there
is no experimental limit on the size of the largest extra
dimension.  However, requiring a large Kaluza-Klein mass gap still
constrained the overall volume of the extra dimensions. Here we
give an explicit model which allows arbitrary large values for
both $V_2$ and $M_{GAP}^2$, at least for the fermions and vectors.
This could have an interesting application in the large extra
dimension scenario because we can have both $\kappa\sim TeV^{-2}$
and small effects from the massive modes by setting a large enough
value of $M_{GAP}$.

In terms of hiding Large Extra Dimensions from our four
dimensional universe, another possible approach is to interpret
all the bulk fields that we have found (massless modes and massive
ones) as a hidden sector, only gravitationally coupled to the
SM. At this level, the SM must then be introduced by
hand, confined on the delta-function brane.  It seems that this is the approach
to take if embedding the Supersymmetric Large Extra Dimenions
scenario in our calculations, proposed in \cite{sled}, to resolve
the Cosmological Constant Problem. This proposal relates the
hierarchy in the Electroweak scale with that of the Cosmological
Constant. The Electroweak scale is set by the size of the extra
dimensions, $r$, and the Cosmological Constant is given by the
Kaluza-Klein mass gap, here fixed by the same scale\footnote{However, the
breakdown of SUSY in the bulk, as in
the solutions studied here, may lead to a larger prediction for
the Cosmological Constant, see \cite{sled}.}
$1/r$. Both may have their
observed values when the 6D fundamental scale is $TeV$, and $r
\sim 0.1mm$. For the mass gap to be this small, SM
particles must be localized to the brane.

Let us end by considering the tunings involved, when constructing
a model with large volume (say, $\sqrt{V_2} \sim 0.1mm$) and large
mass gap (say, $M_{GAP} \sim TeV^{-1}$).  Consider first a large mass gap for
the fermions.  If we set $\alpha \sim 1$ and
$r_1 \ll r_0$,
then the Dirac quantization (\ref{DiracQ}) implies:
\be e\, \frac{r_1}{r_0} \frac{g}{g_1} \sim N
\ee
If we then assume $e \sim 1$ (which is natural from group theory) and $N \sim 1$
(which is required for a small number of families), the large volume - large
mass
gap condition requires a large hierarchy in the bulk gauge
couplings:
\be \frac{g}{g_1} \sim 10^{15} \, .\ee

 Alternatively, we could set $r_1 \sim r_0$ and $\alpha \gg
1$, allowing a large mass gap for both fermions and gauge fields. In this case,
requiring a large mass gap, $M_{GAP} \sim
1/r_0$, as well as large volume, $\sqrt{V_2} \sim \sqrt{\alpha} r_0$,
requires $\alpha \sim 10^{30}$.  Then, again, the Dirac
quantization condition (\ref{DiracQ}) reveals a large hierarchy in
the bulk gauge couplings:
\be \frac{g}{g_1} \sim 10^{-30} \ee
In both scenarios we cannot embed the background
monopole in $U(1)_R$.

There are of course other combinations, for
example with both $\alpha \gg 1$ and $r_1 \ll r_0$, in which these
hierarchies may be relaxed.  However, we should say that these tunings do not
appear to be very natural or promising.  For example, choosing the large
dimensionless number $\alpha \gg 1$ corresponds to heavy negative tension
branes, and deficit angles orders of magnitude less than zero.
On the other hand, independently of trying to embed the Large Extra Dimension
scenario into the present model, we have found an explicit example in which the
Kaluza-Klein mass gap does not go to zero as the volume goes to infinity,
contrary to standard lore.

\section{Conclusions} \label{conclusions}

\begin{table}[top]
\begin{center}
\begin{tabular}{|l|l|}
\hline  Gauge Fields  & $\qquad$ Fermions  \\ \hline
 $V\sim \,\,\,\,({\bf 45},{\bf 1})_0$ & $\lambda \sim \,\,\,\,\,({\bf 45},{\bf
1})_1$ \\
 $\qquad +({\bf 16},{\bf 1})_0$ & $\qquad +({\bf 16}+ \overline{{\bf 16}},{\bf
1})_1$  \\
$\qquad +({\bf 1},{\bf 133})_0$ & $\qquad +({\bf 1},{\bf 133})_1$
\\
 $\qquad +({\bf 1},{\bf 1})_0$ & $\Psi \sim ({\bf 1},{\bf 912})_0$
 \\ \hline
\end{tabular}
\end{center}\caption{\footnotesize The gauge and fermion fields whose
Kaluza-Klein spectrum is given by our work, for the illustrative
example of the anomaly free model $E_6 \times E_7 \times U(1)_R$,
when the monopole is embedded in $E_6$.  We give the quantum numbers
under ${\cal H} = SO(10) \times E_7 \times U(1)_R$, which is the
unbroken subgroup of the 6D gauge group.\label{T:summary}}
\end{table}

In this paper we have analyzed an interesting subsector of gauge
field and fermion fluctuations, in the warped brane world
solutions of 6D minimal gauged supergravities.  In
particular, we have focused on bulk components which could
give rise to SM or Grand Unified gauge and charged
matter fields.

We performed a Fourier
decomposition of 6D fields, and transformed the resulting field
equations into a Schroedinger-like problem.  We were then able to
find the exact solutions for the Kaluza-Klein modes, in terms of
hypergeometric functions. We considered in detail the boundary
conditions that the physical modes must satisfy.  In addition to
the normalizability constraint, consistency also required a
hermiticity condition, which can be interpreted as demanding
current conservation.  We were able to implement this in its general, quadratic
form.  Together, these conditions selected
the physical modes, and gave rise to a discrete mass spectrum,
which we presented in full.  The discreteness of the spectrum is of course to be
expected given the compact topology of the internal manifold,
whose Euler number is two.

Our study can be applied to several sectors of the 6D supergravities.  In Table
\ref{T:summary} we summarise the 6D fields that are covered by our analysis, for
the illustrative example of the anomaly free model $E_6 \times E_7 \times U(1)$.
 Moreover, the corresponding spectra for the non-supersymmetric model of
\cite{Wetterich:1984rv} (at least for the unwarped 4D Poincar\'e invariant case)
and
\cite{carroll},
generalized to Einstein-Yang-Mills with fermions, can be
straightforwardly extracted from those given above simply by setting the warp
factor to one, that is $r_1=r_0$.

The exact results presented in this paper enabled us to study the effects of the
conical
defects, sourced by codimension two branes, on the
Kaluza-Klein wave profiles and mass gaps.  As usual, the gauge fields
have a zero mode with constant wave profile.  For the fermions,
we found that some zero modes can be peaked on a negative tension brane, but in
this case the whole Kaluza-Klein tower is peaked there too.
Therefore, in order to interpret the bulk zero modes as 4D
effective fields of the SM, the mass gap must be
large.

Intriguingly, this does not necessarily drive us to the conventional
Kaluza-Klein picture, with small compact dimensions.  It does not,
because the conical defects allow a novel behaviour in the
mass gap, which can be decoupled from the volume of the
compactification, defined by $\kappa^2/\kappa_4^{\,2}$.  This continues to be
observed in the unwarped limit, where the rugby ball model of \cite{carroll,
sled} is retrieved.  Contrary to standard lore, a finite mass gap can be
obtained, even if the volume goes to infinity.

For example, a
large volume could be arranged in order to generate the Electroweak
hierarchy, whilst maintaining a large mass gap between the
zero modes and the Kaluza-Klein tower. This picture does not seem
to provide a realisation of the Supersymmetric Large Extra Dimension
scenario, where the volume and mass gap should be related.
However, in this way, SM fields could arise from bulk
fields, and along with gravity propagate through the large extra dimensions,
perfectly consistently with observation. Moreover, for better or worse, this
picture
seems to render the Large Extra Dimension scenario less
falsifiable than previously thought, since we do not have to
expect that the bulk Kaluza-Klein modes are accessible at $TeV$
scales.  However, arranging for both a large volume and large mass
gap seems to require a large degree of fine tuning in bulk
couplings. Furthermore, for a more complete idea, we would have to
consider the Kaluza-Klein spectrum for the remaining bulk sectors,
and in particular the gravitational fluctuations to know the
effect of large extra dimensions on post-Newtonian tests.

Indeed, our analysis of 4D effective gauge fields and charged
fermions is only a first step towards a complete analysis of
fluctuations about the warped brane world background in 6D
supergravity.  A final objective would be to derive the full 4D
effective field theory describing light fluctuations, and an
understanding of when 6D physics comes into play.  The sectors
that we have studied here are the simplest ones, in terms of the
mixings, but it should be possible to continue the project to other fields by
extending our work.

Of the remaining sectors to be analyzed, the scalar perturbations
have special importance, since they can contain Higgs fields and can
have implications for the stability of the background solution.
Indeed, in the round sphere limit of the model that we have studied,
tachyons in general emerge from the internal components of the 6D
gauge field orthogonal to the gauge field background
\cite{Randjbar-Daemi:1983bw}.  A first step in the study of the
scalar fluctuations has been presented in \cite{leepapa}, and it
would certainly be interesting if general results can be found.  The
difficulty may be in the mixings of different scalar fluctuations,
which lead to a complicated system of coupled ODEs.

In another direction, much of our analysis was general, and
could also be used to study other theories and other backgrounds with
4D Poincar\'e-2D axial symmetry.

Finally, it would be interesting to investigate whether there exist other
mechanisms, which lead to the same decoupling between the mass gap and internal
volume that we have found here.  Indeed, in our set up we have been able to show
that the decoupling arises due to the conical defects, but it may be possible to
find other sources in different frameworks.  In this way, our explicit example
may be a realisation of a more general mechanism.

\vspace{1cm}

{\bf Acknowledgments.}   It is our pleasure to thank Giulio Bonelli, Luca
Ferretti, Cristiano Germani, Tony Gherghetta, Sean Hartnoll, Martin O'Loughlin,
Giuliano Panico and Roberto Valandro for interesting conversations.

\newpage

\appendix

{\Large \bf Appendix}

\section{Conventions and Notation} \label{conventions}\setcounter{equation}{0}

We choose the signature $-,+,+,+,...$ for the metric $G_{MN}$. The
Riemann tensor is defined as follows
\be R_{MNS}^{R}=\partial_M \Gamma_{NS}^R -\partial_N \Gamma_{MS}^R
+ \Gamma_{MP}^R \Gamma_{NS}^P -\Gamma_{NP}^R \Gamma_{MS}^P,  \ee
where the $\Gamma 's$ are the Levi-Civita connection. Whereas the
Ricci tensor and the Ricci scalar
\be R_{MN}=R_{PMN}^{P}, \ \ \ \ R=G^{MN}R_{MN}.  \ee
Here $M,N,...$ run over all space-time dimensions.

Our choice for the 6D constant gamma matrices $\Gamma^A$,
$A=0,1,2,3,5,6,$ is
\be \Gamma^{\mu}=\left(\bacc 0 & \gamma^{\mu} \\
\gamma^{\mu} & 0 \ea \right), \quad  \Gamma^5=\left(\bacc 0 & \gamma^5 \\
\gamma^5 & 0 \ea \right),\quad  \Gamma^6=\left(\bacc 0 & -i \\
i & 0 \ea \right), \label{Gamma}\ee
where the $\gamma^{\mu}$ are the 4D constant gamma matrices and
$\gamma^5$ the 4D chirality matrix. Moreover the spin connection
is
\be
\Omega^{[A,B]}_M=\eta^{BC}\Omega_{M\,\,\,C}^{\,\,A}=\eta^{BC}\left(e^A_N\Gamma_{
MR}^Ne^R_B
+e^A_N\partial_Me^N_B\right),\ee
where $e^A_M$ is the vielbein.

\section{Delta-Function Singularities} \label{deltafn}\setcounter{equation}{0}

In this appendix we briefly review how the Ricci scalar acquires a
delta-function contribution in the presence of a deficit angle, and
examine what this implies for our choice of metric function $e^B$
in (\ref{axisymmetric}).

Let us consider the ansatz (\ref{axisymmetric},\ref{assumeeB}), and
illustrate the case for the deficit angle $\delta$ at $\rho=0$. Near
 $\rho=0$ the metric $ds_2^2$ of the 2D internal space can be
written as follows
\be
ds^2_2=d\rho^2+\left(1-\frac{\delta}{2\pi}\right)^2\rho^2d\varphi^2 \, .\ee
By using the change of coordinate
$r^{1-\delta/2\pi}/(1-\delta/2\pi)=\rho$, this metric becomes
\be
ds^2=r^{-\delta/\pi}\left(dr^2+r^2d\varphi^2\right).\label{s2}\ee
From (\ref{s2}) one can show \cite{Randjbar-Daemi:2004ni}
\be
R=2\, \delta \, r^{\delta/\pi}
\delta^{(2)}\left(\bf{y}\right)+...,\label{Rdelta}\ee
where the 2D vector ${\bf y}$ is defined by ${\bf y} = (r \cos
\varphi, r \sin \varphi)$, $\delta^{(2)}$ is the 2D Dirac
delta-function and the dots are the smooth
contributions\footnote{Eq. (\ref{Rdelta}) describes the asymptotic behaviour in
the vicinity of the
brane \cite{Randjbar-Daemi:2004ni}.  It seems that for positive
$\delta$ the Ricci scalar vanishes at the
origin. 
 However, the first term on the
right hand side of (B.3) should be interpreted in a distributional
sense. The effect of the Ricci scalar as a distribution on a scalar
test function $f({\bf y})$ is then $\int d^2{\bf y} \sqrt{g} R
f({\bf y}) = 2 \, \delta f({\bf 0}) + \dots$.}. On the other hand, near to
$\rho=0$ the Ricci scalar, $R$, can be expressed in terms of
derivatives of $B$:
\be R=-B''-\frac{1}{2}(B')^2, \label{RB}\ee
where $' \equiv \partial_{\rho}$, and from (\ref{Rdelta}) and (\ref{RB}) it
follows that
\be
B''=-2\, \delta \, r^{\delta/\pi}
\delta^{(2)}\left(\bf{y}\right)+...,\label{Bdelta}\ee
That is, the metric function $e^B$ contains a delta-function
contribution in its second order derivative with respect to $\rho$. In
coordinate system
(\ref{u}), (\ref{Bdelta}) becomes
\be \partial^2_u
B=-2\, \delta \, r^{\delta/\pi}
\delta^{(2)}\left(\bf{y}\right)+...,\label{B2u}\ee

The delta-function in the curvature also gives rise to a delta-function in the
derivative of the spin connection (\ref{spinconnection}).  The Riemann tensor is
defined in terms of the spin connection as:
\be
 R_{\,\,\,\,B}^{\,\,A}=d\Omega_{\,\,\,\,B}^{\,\,A}+\Omega_{\,\,\,\,C}^{\,\,A}
\wedge\Omega_{\,\,\,\,B}^{\,\,C} \,.
\label{Riemann-conn}\ee
Near to the brane $\rho=0$, relation (\ref{Riemann-conn}) gives
$R_{\rho
\varphi\,\,\,6}^{\,\,\,5}=\partial_{\rho}\Omega_{\varphi\,\,\,6}^{\,\,5}$, and
from the expression for the spin-connection (\ref{spinconnection}):
\be \Omega'=-\left(\frac{1}{2}B''+\frac{1}{4}(B')^2\right)e^{B/2},
\label{Oprime}\ee
thus leading to the delta-function behaviour from (\ref{Bdelta}).

These results must be recalled when obtaining the
Schroedinger-like equations that govern the fluctuations.

\section{Imposing Boundary Conditions }
\label{boundcond}\setcounter{equation}{0}

Here we study the implications of the NC and the HC for gauge
field fluctuations.  To impose the boundary conditions the following
properties will be useful:
\bea &&\quad F(a,b,c,z)\stackrel{z\rightarrow0}{\rightarrow}1,\label{z1} \\
&&F(a,b,c,z)=\Gamma_1 F(a,b,a+b-c+1,1-z)\nonumber \\
&&+\Gamma_2 (1-z)^{c-a-b}F(c-a,c-b,c-a-b+1,1-z),\label{1-z} \eea
where
\be \Gamma_1\equiv
\frac{\Gamma(c)\Gamma(c-a-b)}{\Gamma(c-a)\Gamma(c-b)}, \quad
\Gamma_2\equiv \frac{\Gamma(c)\Gamma(-c+a+b)}{\Gamma(a)\Gamma(b)},
\label{G11}\ee
and $\Gamma$ is the Euler gamma function. The relation (\ref{1-z})
is valid if $c-a-b$ is not an integer \cite{whittaker} and $y_1$
and $y_2$ in (\ref{y12}) are both well defined when $c$ is not an
integer. In general $c-a-b$ and $c$ are not integers for generic
$\omega$ and $\ob$; so we can consider $\omega$ and $\ob$ as
regulators to use (\ref{y12}) and (\ref{1-z}) and at the end we
can take the limits in which $c-a-b$ and $c$ go to an integer,
which will turn out to be well defined.

We first consider the behaviour of $\psi$ for $u\rightarrow
\overline{u}$, that is $z\rightarrow 0$ because of the definition
$z=\cos^2\left(\frac{u}{r_0}\right)$. For $c\neq 1$ we use the
expression for $\psi$ given in (\ref{cnot1}) and property
(\ref{z1}) gives us
\be \psi \stackrel{u\rightarrow
\overline{u}}{\rightarrow}K_1(\overline{u}-u)^{2\gamma}+K_2(\overline{u}-u)^{
1-2\gamma},\label{uub}\ee
where we used $c=1/2+2\gamma$. So the NC (\ref{NCV}) implies $K_1=0$
when $\gamma\leq -1/4$ and $K_2=0$ when $\gamma\geq 3/4$. On the
other hand the HC (\ref{HCV2}) implies\footnote{In fact, for the gauge field
sector, each term in the HC (\ref{HCV2}) is separately zero if one requires
their finiteness.  Therefore (\ref{HCa}) and its counter-part for $u=0$ are
sufficient to ensure (\ref{HCV2}).  On the other hand, for the fermions, there
are some cases in which they are each finite and non-zero, and so (\ref{HCV2})
requires that they cancel.}
\be \lim_{u\rightarrow
\overline{u}}\psi^*\left(-\partial_u+\frac{1}{2}\frac{1}{u-\overline{u}}
\right)\psi<\infty\label{HCa}\ee
and by using the behaviour (\ref{uub}) this limit becomes
\be \left(2\gamma -\frac{1}{2}\right)\lim_{u\rightarrow
\overline{u}}\left[|K_1|^2(\overline{u}-u)^{4\gamma-1}
-K_1^*K_2+K_1K_2^*-|K_2|^2(\overline{u}-u)^{-4\gamma+1}\right],\ee
so the HC implies $K_1=0$ when $\gamma <1/4$ and $K_2=0$ when
$\gamma >1/4$. The case $\gamma=1/4$ corresponds to $c=1$ and so we
have to use the expression of $\psi$ given in (\ref{wronskian}).
We have then
\be \psi \stackrel{u\rightarrow
\overline{u}}{\rightarrow}K_1(\overline{u}-u)^{1/2}-K_2(\overline{u}-u)^{1/2}
\ln(\overline{u}-u)\ee
for which (\ref{HCa}) implies $K_2=0$. Therefore we obtain (\ref{psi1}) and
(\ref{psi2}).

The discreteness of the spectrum emerges when we impose the NC and
HC for $u\rightarrow 0$. For instance for $m\geq N_V$, up to an
overall constant, the behaviour of $\psi$ is given by properties
(\ref{z1}) and (\ref{1-z}):
\be  \psi \stackrel{u\rightarrow
0}{\rightarrow}\Gamma_1u^{2\b}+\Gamma_2u^{1-2\b},\label{psib}\ee
where $\Gamma_{1,2}$ are defined in (\ref{G11}) and we used
$c-a-b=1/2-2\b$. Behaviour (\ref{psib}) is similar to (\ref{uub})
but $\gamma$ is replaced by $\beta$. So, following the same steps as above, the
NC and the HC imply
that $\Gamma_1=0$ for $\beta<1/4$ and $\Gamma_2=0$ for $\beta>1/4$. Let
us study the case $m\geq N_V$ and $\b<1/4$, that is
\be N_V\leq m < 0.\label{assumem}\ee
 We then have
\be 0=\Gamma_1\equiv
\frac{\Gamma(c)\Gamma(c-a-b)}{\Gamma(c-a)\Gamma(c-b)}.\ee
Since the Euler gamma function never vanishes we require that
$\Gamma(c-a)=\infty$ or $\Gamma(c-b)=\infty$ and this is possible
only when $c-a=-n$ or $c-b=-n$, where $n=0,1,2,3,...$. By using
the definitions (\ref{gbetaabcV}) both conditions lead to the following squared
masses
\be
M_{V\,n,m}^2=\frac{4}{r_0^2}\left\{n(n+1)+\left(\frac{1}{2}+n\right)\left[
-m\omega+(m-N_V)\ob\right]\right\}\ee
which are positive because of (\ref{assumem}) and $n\geq 0$. When
$m\geq N_V$ and\footnote{The $\b =1/4$ case is recovered by taking
the limit $\omega \rightarrow 0$.} $\beta\geq 1/4$, that is
\be m\geq N_V \quad and \quad  m \geq 0,\label{assumem2}\ee
we have
\be 0=\Gamma_2\equiv
\frac{\Gamma(c)\Gamma(-c+a+b)}{\Gamma(a)\Gamma(b)}\ee
and this implies $a=-n$ or $b=-n$, where $n=0,1,2,3,...$. The
corresponding squared masses are
\be
M_{V\,n,m}^2=\frac{4}{r_0^2}\left\{n(n+1)+\left(\frac{1}{2}+n\right)\left[
m\omega+(m-N_V)\ob\right]+m\omega(m-N_V)\ob\right\}\ee
which are positive or vanishing. We can study the case $m<N_V$ in
a similar way. The complete result for the gauge fields sector is
given in equations (\ref {MV1})-(\ref{MV4}). We observe that
(\ref{HCV2}) is now automatically satisfied by every pair of wave
functions $\psi$ and $\psi'$, for a given quantum number $m$,
since the asymptotic behaviour of the wave function cannot depend
on the quantum number $n$: this is a consequence of the $1/u^2$
and $1/(\overline{u}-u)^2$ singularities of the potential $V$ in
(\ref{SchroedingerVGGP}).

\section{Complete Fermionic Mass Spectrum}
\label{fermspectrum}\setcounter{equation}{0}
In this appendix we give the complete fermionic spectrum which is
also labeled by an integer quantum number $n=0,1,2,...$. Although
much longer, the calculation proceeds in exactly the same way as
for the gauge field sector, outlined in the previous
appendix\footnote{There is one additional subtlety.  Here, for
the values of $m$ which allow a zero mode, we must impose a mixed
HC between the massless mode and massive modes, in addition to the
diagonal HC. In general the HC involving distinct wave functions
$\psi_{m n}$ and $\psi_{m n'}$ does not lead to additional
constraints, because the asymptotic behaviour of the modes is
independent of $n$. However, the massless modes are more strongly
constrained than the massive ones, obeying as they do a decoupled
Dirac equation in addition to the Schroedinger equation.}.

\begin{description}

\item[{\Large $m \geq -\frac{1}{2}+N+\frac{1}{2\ob}$}]: in this
case $K_2=0$ and we obtain the following squared masses.

\begin{itemize}
\item For $m>-\frac{1}{2}+\frac{1}{2\omega}$
\be
M^2_{F\,n,m}=\frac{4}{r_0^2}\left[\frac{1}{2}+n+\omega\left(\frac{1}{2}
+m\right)\right]
\left[\frac{1}{2}+n +\ob\left(\frac{1}{2}+m-N\right)\right]>0.
\label{m1} \ee
\item For
$-\frac{1}{2}-\frac{1}{2\omega}<m<-\frac{1}{2}+\frac{1}{2\omega}$
\be
M^2_{F\,n,m}=\frac{4}{r_0^2}\left[\frac{1}{2}+n+\omega\left(\frac{1}{2}
+m\right)\right]
\left[\frac{1}{2}+n +\ob\left(\frac{1}{2}+m-N\right)\right]>0. \label{m2} \ee
or
\be
M^2_{F\,n,m}=\frac{4}{r_0^2}(1+n)\left[1+n+\ob\left(\frac{1}{2}
+m-N\right)-\omega\left(\frac{1}{2}+m\right)\right]>0.\ee
\item For $m\leq -\frac{1}{2}-\frac{1}{2\omega}$
\be
M^2_{F\,n,m}=\frac{4}{r_0^2}(1+n)\left[1+n+\ob\left(\frac{1}{2}
+m-N\right)-\omega\left(\frac{1}{2}+m\right)\right]>0.
\label{m4}\ee
\end{itemize}
\item[{\Large $m \leq -\frac{1}{2}+N-\frac{1}{2\ob}$}]: in this
case $K_1=0$ and we obtain the following squared masses.

\begin{itemize}
\item For $m>-\frac{1}{2}+\frac{1}{2\omega}$
\be
M^2_{F\,n,m}=\frac{4}{r_0^2}n\left[n-\ob\left(\frac{1}{2}+m-N\right)
+\omega\left(\frac{1}{2}+m\right)\right]\geq 0.  \ee
\item For
$-\frac{1}{2}-\frac{1}{2\omega}<m<-\frac{1}{2}+\frac{1}{2\omega}$
\be
M^2_{F\,n,m}=\frac{4}{r_0^2}n\left[n-\ob\left(\frac{1}{2}+m-N\right)
+\omega\left(\frac{1}{2}+m\right)\right]\geq 0. \label{02} \ee
\item For $m\leq -\frac{1}{2}-\frac{1}{2\omega}$
\be
M^2_{F\,n,m}=\frac{4}{r_0^2}\left[\frac{1}{2}+n-\omega\left(\frac{1}{2}
+m\right)\right]\left[
 \frac{1}{2}+n-\ob\left(\frac{1}{2}+m-N\right)\right]>0. \ee
\end{itemize}

\item[{\Large $-\frac{1}{2}+N-\frac{1}{2\ob}<m <
-\frac{1}{2}+N+\frac{1}{2\ob}$}]: this case is possible only when
$\overline{\delta}<0$.

\begin{itemize}
\item For $m>-\frac{1}{2}+\frac{1}{2\omega}$ we have $K_1=0$ and
\be
M^2_{F\,n,m}=\frac{4}{r_0^2}n\left[n-\ob\left(\frac{1}{2}+m-N\right)
+\omega\left(\frac{1}{2}+m\right)\right]\geq 0.\ee
\item For $m\leq -\frac{1}{2}-\frac{1}{2\omega}$ we have two
possibilities. We have $K_2=0$ and
\be
M^2_{F\,n,m}=\frac{4}{r_0^2}(1+n)\left[1+n+\ob\left(\frac{1}{2}
+m-N\right)-\omega\left(\frac{1}{2}+m\right)\right]>0\ee
or $K_1=0$ and
\be
M^2_{F\,n,m}=\frac{4}{r_0^2}\left[\frac{1}{2}+n-\omega\left(\frac{1}{2}
+m\right)\right]\left[
 \frac{1}{2}+n-\ob\left(\frac{1}{2}+m-N\right)\right]>0\ee
\item $-\frac{1}{2}-\frac{1}{2\omega}<m<-\frac{1}{2}+\frac{1}{2\omega}$:
this case is possible only when $\delta<0$ and we obtain $K_1=0$ and
\be
M^2_{F\,n,m}=\frac{4}{r_0^2}n\left[n-\ob\left(\frac{1}{2}+m-N\right)
+\omega\left(\frac{1}{2}+m\right)\right]\geq 0. \label{mf}\ee
\end{itemize}
\end{description}

Again, we can perform a check of our results by considering the
$S^2$ limit ($\omega\rightarrow 1$, $\ob\rightarrow 1$).  In this
case, the mass spectrum (\ref{m1})-(\ref{mf}) reduces correctly to
\be
a^2M^2_F=\left(l+\frac{1+N}{2}\right)\left(l+\frac{1-N}{2}\right),\quad
multiplicity=2l+1\ee
where\footnote{The number $l$ is defined in different ways in
equations (\ref{m1})-(\ref{mf}). For instance we have $l\equiv
n+m+(1-N)/2$ for (\ref{m1}) and $l\equiv 1/2+n-N/2$ for
(\ref{m4}).} $l=\frac{|N|-1}{2}+k$ and $k=0,1,2,3,...$.

\end{document}